
\magnification=\magstep1

\hsize=6truein
\vsize=8.5truein
\raggedbottom
\nopagenumbers
\tolerance=10000

\font\bfs=cmbx7

\font\bigbf=cmbx10 scaled \magstep1
\font\bigbold=cmbx10 scaled \magstep2

\def\hang{\parshape
     2 0in 6.00truein .20truein 5.80truein
   \noindent}

\def\skip{\vskip.15truein}

\def\today{\ifcase\month\or January\or February\or March\or April\or May\or
June\or July\or August\or September\or October\or November\or December\fi
   \space\number\day, \number\year}

\def\pmb#1{\setbox0=\hbox{#1}%
     \kern-.025em\copy0\kern-\wd0
     \kern.05em\copy0\kern-\wd0
     \kern-.025em\raise.0433em\box0 }

\def\1{$^{-1}$}

\nopagenumbers

\headline={\ifnum\pageno=1 \hss{\bigbf } \hss
  \else{\bfs Anderson, Neumann \& Perelson\hfil page \folio}\fi}
 \voffset=2\baselineskip

\hbox{ }
\vskip-20pt
\centerline {\bigbold A Cayley Tree Immune Network Model}
\smallskip
\centerline {\bigbold with Antibody Dynamics}
\skip
\centerline {Russell W. Anderson}
\centerline {Theoretical Biology and Biophysics}
\centerline {Mail Stop K710}
\centerline {Los Alamos National Laboratory}
\centerline {Los Alamos, NM  87545}
\centerline {email: rwa@temin.lanl.gov}
\centerline {office: (505) 667-9455}
\centerline {FAX:(505) 665-3493}
\smallskip
\centerline {Avidan U. Neumann}
\centerline {Santa Fe Institute}
\centerline {1660 Old Pecos Trail}
\centerline {Santa Fe, NM 87501}
\centerline {email: aun@santafe.edu}
\smallskip
\centerline {Alan S. Perelson}
\centerline {Theoretical Biology and Biophysics}
\centerline {Los Alamos National Laboratory}
\centerline {Los Alamos, NM  87545}
\centerline {email: asp@receptor.lanl.gov}
\vskip.2truein
\baselineskip=15pt
\def\fhi{f(h_i)}

\centerline {For copies of the figures,}
\centerline {please write or FAX R. Anderson}
\skip
\leftline{\bigbf ABSTRACT}
\skip
A Cayley tree model of idiotypic networks that includes
both B cell and antibody dynamics is formulated and analyzed.
As in models with B cells only, localized states
exist in the network with limited numbers of activated clones
surrounded by virgin or near-virgin clones. The existence
and stability of these localized network states are explored
as a function of model parameters.
As in previous models that have included antibody,
the stability of immune and tolerant
localized states are shown to depend on the ratio
of antibody to B cell lifetimes as
well as the rate of antibody complex removal.
As model parameters are varied, localized steady-states can
break down via two routes: dynamically, into chaotic attractors, or
structurally into percolation attractors.
For a given set of parameters, percolation and chaotic
attractors can coexist with localized attractors,
and thus there do not exist clear cut boundaries in parameter
space that separate regions of localized attractors from
regions of percolation and chaotic attractors.
Stable limit cycles, which are frequent in the
two-clone antibody B cell (AB) model, are only observed
in highly connected networks. Also found in highly
connected networks are localized chaotic attractors.
As in experiments by Lundkvist et al. (1989), injection
of $Ab_1$ antibodies into a system operating in the chaotic
regime can cause a cessation of fluctuations of $Ab_1$
and $Ab_2$ antibodies, a phenomenon already observed in
the two-clone  AB model.
Interestingly, chaotic fluctuations continue at higher levels of
the tree, a phenomenon observed by Lundkvist et al. but not
accounted for previously.

\vfil\eject\skip
\leftline{\bigbf 1. INTRODUCTION}
\medskip
Jerne (1974) postulated that the immune system
functions as a network, where lymphocytes
are stimulated or suppressed by ``idiotypic" interactions
with complementary antibodies and immunoglobulin receptors.
Since then, experimental evidence of an active
immune network has been found
(Holmberg et al., 1984; Kearney and Vakil, 1986;
Lundkvist et al., 1989).
Several theories have been advanced for a biological
function of this idiotypic network,
among them is the idea that immunological
memory is a dynamic consequence of network interactions
(Hoffmann, 1975; Richter, 1975; Farmer et al., 1986;
Weisbuch, 1990; Weisbuch et al., 1990; Behn et al., 1992).
Under the ``dynamic memory hypothesis"
after initial antigen exposure,
an expanded, neutralizing clonal population
is sustained through network interactions with
idiotypically related clones.
Mathematical models have been formulated to make these
ideas more precise (for reviews see
Perelson, 1989;
Varela and Coutinho, 1991;
and De Boer et al., 1992a).

\skip
Immune network models can be
classified by the degree of complexity with
which they model
(i) the structure of network connectivity
and (ii) the dynamics of individual clonal species.
Network structure has been modeled with varying
degrees of realism.

\skip
The simplest model structure describes
the dynamics of a pair of complementary B cell clones.
We refer to this class as ``two-clone models".
Variations of these models have been studied extensively
(Perelson, 1989; De Boer et al., 1990;
Stewart and Varela, 1990).
(A rigorous dynamical analysis of two-clone models -
under a variety of assumptions for clonal dynamics - is given
by De Boer, Kevrekidis and Perelson (1993a,b).)
Two-clone models have the advantage of mathematical tractability
and shed light on the {\it dynamics} of clonal
populations as a function of model assumptions
and parameters. But, they are
insufficient for investigations of
the effects of network structure on dynamical behavior.

\skip
The next level of model complexity introduces
network connectivity.
This has been done in two ways.
One method prescribes the
network structure using static, {\it a priori}
connectivity assumptions. For example,
Cayley tree models assume a uniform connectivity structure
(Weisbuch et al., 1990).
Other models prescribe a network structure
based on experimentally known interactions
(Stewart and Varela, 1989) or assume random connectivity
matrices (Hoffmann, 1982; Spouge, 1986; De Boer, 1988).
The second method allows network connectivity to
develop from assumptions of
affinity matching rules, which in turn,
determine idiotypic interactions.
This class of models includes bit-string models
(Farmer et al., 1986; De Boer and Perelson, 1991;
Celada and Seiden, 1992),
as well as other shape-space models
(Segel and Perelson, 1988, 1989; Weinand, 1990;
Weisbuch, 1990;
Stewart and Varela, 1991; De Boer et al., 1992b).
\skip
In this study, we analyze one type of prescribed
network model: a homogeneous Cayley tree model.
Previous models of this class
modeled B cell populations but not their corresponding
antibodies
(Weisbuch et al., 1990; Neumann and Weisbuch, 1992a,b).
It was shown that in certain parameter ranges,
this model possesses localized steady-states,
where a large population at one level could be
sustained by idiotypic interactions with
small or intermediate populations of clones at the next level,
and neighboring clones in the network would
remain virgin or near-virgin.
For example, if
antigen is assumed to only interact with the level
1 clone, then a localized state occurs
when the second level populations are not high enough to
stimulate proliferation of third level populations.
These localized states were presented as models
of immune network ``memory" or ``tolerance",
depending upon whether the field at level 1
was low or high.
\skip
In their analysis of two-clone models, De Boer et al. (1993a,b)
show that when antibody dynamics
are included,
what would be stable system attractors in a simple B cell
model may become oscillatory or chaotic,
depending on parameter values.
Here, we
investigate the effect of antibody dynamics
on the stability of states in a Cayley tree
model.
We call this model the ``AB Tree model",
where AB stands for antibody and B cell dynamics,
and Tree stands for the Cayley tree topology.
\skip
We show that
the addition of antibody dynamics does not
substantially alter the steady-states observed
in the work of Weisbuch and colleagues
and that isolated, non-oscillatory states are readily
obtained.
We derive conditions for the existence and stability of
these localized states and perform bifurcation
analyses on the model.
As network connectivity is increased, localized steady-states
disappear and only chaotic attractors and
nonlocalized steady-states remain.
Non-localized steady-states are referred to as ``percolation"
attractors. They are states in which alternating
levels of clones are activated throughout
the network.
As dynamical parameters (such as the ratio of
the antibody death rate
or complex removal rate to the B cell death rate) are decreased,
stable localized  steady-states lose stability and
and trajectories approach what appear to be chaotic attractors.
These attractors, as is the case with percolation
attractors, generally do not remain
localized in the network, and signals
eventually propagate through successive levels of the
network.  Information about initial conditions
is generally lost in both of these attractors and
there is no way to know which level was originally stimulated.
However, we do find localized chaotic attractors
and limit cycle behavior in parameter regimes characterized
by high connectivity.
Nevertheless, if real immune systems operate in parameter domains
characterized by nonlocalized behavior, it would be
difficult to see how they could account for dynamical memory.
Finally, we will discuss other limitations of this approach
toward understanding immune network behavior.
\skip
\leftline{\bigbf 2. THE AB Tree MODEL}
\medskip

We consider individual B cell clones which are formed in
the bone marrow, proliferate in
response to stimulation, and die in the periphery.
The corresponding antibodies are secreted by the B cells
in response to stimulation, decay in the periphery,
and are actively removed or inactivated by complex
formation with other antibodies.
Following previous work
(Weisbuch et al., 1990; Varela and Coutinho, 1991; Perelson,
1989; De Boer and Perelson, 1991),
we assume for each clone $i$,
the total amount of idiotypic stimulation is a linear combination
of the concentration of
antibodies of all other clones $j$.
The amount of stimulation detected by a clone $i$ is referred
to as it's {\it field}, $h_i$:
$$ h_i = \sum_j J_{ij}a_j \ \ ,\eqno(2.1) $$
where $J_{ij}$ is the affinity between clone $i$ and the
antibodies of clones $j$ and $a_j$ is the concentration
of antibody $j$.  We assume
$J_{ij} = 0$ (no interaction) or
$J_{ij} = 1$ (maximum affinity).
Without loss of generality one can make the maximum
affinity any real positive number, $K$, rather than 1,
however for reasons of simplicity we choose $K=1$.

Network structure is determined by this affinity, or
connectivity matrix.
\skip
Proliferation of B cell clones and antibody secretion rates
are a function of their stimulation.
In this model, we use
a phenomenological biphasic activation function:
$$
f(h_i) =
	{h_i\over(\theta_1+h_i)}
	 \
	{\theta_2\over (\theta_2+h_i)}\ \ .\eqno(2.2) $$
where $\theta_2 \gg \theta_1$.
The maximum activation level is close to 1
and occurs at the intermediate field strength,
$h=\sqrt{\theta_1 \theta_2}$.
The use of this function has theoretical and experimental
justification and has been used extensively in immune
system models (Varela and Coutinho, 1991; Perelson, 1989;
De Boer and Perelson, 1991; De Boer et al., 1992a,b, 1993a,b).
Activation is thought to be proportional to the
proportion of surface immunoglobulin that is
crosslinked.
Biophysical models of receptor crosslinking of
bivalent ligands predict
a symmetric, log bell-shaped crosslinking curve
(Perelson and DeLisi, 1980); furthermore,
antibody production follows a similar empirical dose-response
curve (Celada, 1971).
\skip
We model the population change of clone $i$ with a
pair of differential equations representing the B cells
$b_i$ and the concentration of their antibodies $a_i$:
$$
\eqalign{
{db_i\over dt} &= m + pf(h_i)b_i-d_Bb_i\ \ ,\cr
\noalign{\vskip.1truein}
{da_i\over dt} &= sf(h_i)b_i -d_Aa_i-d_Ca_ih_i\ \ ,\cr
}
\eqno(2.3) $$
where $f$ is the activation function,
$m$ is the bone marrow source rate, and
$d_B$ is the B cell death rate.
The proliferation parameter, $p$, must be such that
when B cells are stimulated, their growth rate exceeds their
death rate or else no clonal expansion would
occur; thus,
$$p  > d_B \  \ . \eqno (2.4) $$
Parameters in the antibody equations are $s$,
the secretion rate,
$d_A$, the antibody decay rate, and
$d_C$, the rate of complex formation and removal.
The parameter $d_C$
is a combination of several physical parameters, e.g.
$d_C=\hat d_c v^2 K$, where
$\hat d_c$ is the rate of complex elimination by macrophages
and phagocytic cells, $v$ is the valence of the antibody,
and $K$ is the affinity of the idiotype for
anti-idiotypic antibodies
(De Boer et al., 1993a,b).

\skip
In two-clone models, the fields $h_1$ and $h_2$ are
simply the complementary antibody
populations $a_2$ and $a_1$, respectively.
In the
Cayley-tree model, the fields incorporate a branching
network structure where each clone is connected to
$z$ other clones (see Fig. 1). The parameter
$z$ is called the coordination number of the Cayley tree.
The field for the root, or
first level clone, is then

$$h_1 = za_2\ \ , \eqno (2.5)$$
since the clone at level 1 interacts with $z$ clones
at level 2.
If antigen is present, it is assumed to react only
with the clone at level 1, and the field becomes
$$h_1 = za_2 + Ag\ \ , \eqno (2.6)$$
where $Ag$ represents the effective antigen concentration
(the actual antigen concentration, multiplied by
it's valence and affinity).
In fact, this property of antigen reactivity
defines level 1.
Note that in this model,
all antibodies at a given level are treated
equivalently.
The state variables $b_i$ and $a_i$ thus represent
a single B cell or antibody population, which is the
same for all populations at a given level.
All subsequent clones experience a field:

$$h_i = a_{i-1} + (z-1)a_{i+1}	\ \ , \ \ i>1\ \ ,
\ \ z \geq 2 \ .  \eqno(2.7)
$$

\skip
\leftline{\bigbf 2.1 Parameter Values}
\skip
Previous modeling studies have provided estimates
for the model parameters
(Varela and Coutinho, 1991; De Boer and Perelson, 1991;
De Boer et al., 1993a,b).
Briefly, typical parameter estimates are as follows:
Due to cell division at the pre-B cell stage,
each clone will consist of approximately 10-20 cells
when it is generated.
Here, we assume that the bone marrow produces cells of
clone $i$ at a constant rate $m$.  Because the same clones are
probably not produced every day, we use as an average
production rate about one cell per clone per day, $m\approx 1$.
B cells have a lifetime of about 2 days,
$d_B \approx 0.5$~d\1.
Activated cells divide about every 16 hours,
$p \approx 1$~d\1.
Antibodies may persist much longer,
about 20 days; thus,
$d_A \approx 0.05$~d\1.
(Varela and Coutinho estimate $d_A \approx 0.1$~d\1.)
A unit of antibody is the amount of antibody produced by
a fully matured B cell in one day, thus, measured in
units, $s = 1$~d\1.
Antibody complexes are removed at a rate
$d_C \approx 10^{-2}$~d\1 ~unit\1,
estimated in De Boer and Perelson (1991)
where the notation
$d_C = d_cK$ was used.
The threshold for proliferation is set at
$\theta_1=100$.
The onset of suppression, or the higher threshold of the
dose-response curve, is generally set several orders
of magnitude higher, i.e. $\theta_2=10^4$.
Throughout this paper, these estimated parameters
will be referred to as the ``standard" parameter set.
\skip
Connectivity can be defined and
measured in a number of ways.
In a young mouse, any given antibody will
crossreact with as much as 23-28\% of the other
antibodies; in adult mice,
this percentage reduces to about 1-2\%
(Holmberg et al., 1984; Kearney et al., 1987).
However, affinities of IgM molecules in immature
immune systems are relatively low and nonspecific.
Using accessibility computations,
Novotn\'y et al., (1987) estimate that a
immunoglobulin molecule has 40 distinct idiotypic
determinants available for anti-idiotypic binding.
Not all of these epitopes
will necessarily
result in an idiotypic response, while
more than one antibody may bind to others.
Thus, although 40 could serve as a reasonable
estimate for $z$, for our standard parameter
set, we choose a more conservative, intermediate value
of $z=10$.

\skip
\leftline{\bigbf 3. STEADY-STATES}
\skip
In the following analysis, we find the conditions for the
existence of localized states for the AB Tree model.
We derive estimates for the
steady-state B cell populations and their corresponding
antibodies at each level in the network.
We then apply stability analysis to these
steady-states to find the conditions for
the {\it stability} of these localized states.

In previous analyses of models that contain only B cells
(B models) and that employ a log bell-shaped
activation function
(Weisbuch et al., 1990; De Boer et al., 1990, 1992a),
three possible equilibrium levels
for each B cell population have been identified:
\skip
1.) a virgin, or unstimulated, level, $m/d_B$ ,
\skip
2.) a large population level
corresponding to cells in an ``immune" state,
that experience a low activating field,
${d_B \over {(p-d_B)}} \theta_1$ (see Eq. 3.7),
and
\skip
3.) an intermediate population level
corresponding to cells in a ``suppressed" state,
that experience a high suppressive field,
${(p-d_B) \over d_B} \theta_2$ (see Eq. 3.5).
\skip
To a good approximation, localized network attractors
consist of B cell populations at these various levels
(Weisbuch et al., 1990).
The purpose of this study is to investigate the behavior
of the Cayley tree model when antibody dynamics are introduced;
therefore, it is of interest to examine how the system attractors
are affected by the introduction of dynamical equations for
the antibodies.
\skip
When B cell populations are unstimulated, (i.e.
$f(h_i)=0$ $\forall i$),
all B cell populations attain
the virgin steady-state; while the corresponding antibody
populations diminish to zero.
This is an important, but dynamically uninteresting,
system attractor corresponding to the resting
state of a classical clonal selection
immune response model.
It would correspond to
a completely decoupled immune network.
In our network model, this state is not attainable
if even one antibody population is non-zero at steady-state.
As we shall see, however, a near-virgin steady-state is
possible under some conditions.
\skip
Other system attractors likely to exist in this model are
localized memory and percolation attractors.
A localized memory state occurs when the clonal population at one
level is high
while all other levels remain suppressed or at near-virgin levels.
For example, a localized memory at level 1 corresponds to a
high, activated population of level 1 clones, sustained by
an intermediate, suppressed population of level 2 clones.
In order for this state to be considered localized,
levels 3 and beyond must remain at low, or near-virgin levels.
This attractor is called a localized memory because the
antigen-reactive clone at level 1 is high and capable
of quickly eliminating antigen as in a typical
secondary immune response.
(In a percolation attractor, by comparison, levels 3
and beyond would experience
activating fields.)

\skip
The first localized state of interest is a special case.
In this state level 1 is activated, level 2 suppressed,
and all others near-virgin. Other localized states (where a level
other than the first is activated) are a generalization
of this result, since connectivity backward through the network
must be taken into account. The conditions for localized memory
will be shown to be only slightly more restrictive in the general
case (see Section 5).
\skip
Estimates for the steady-state populations are greatly simplified
using the following approximations:
{}From Eq. (2.3), for small $m$
(bone marrow source term), approximate B cell
equilibria are obtained at the intersections of
the curve $y=pf(h)$ with the line $y=d_B$ (the B cell
death rate), i.e. at

$$p\fhi \approx d_B\ \ . \eqno (3.1)$$
Since we are seeking a steady-state with
level 2 suppressed, we assume $h_2\gg\theta_1$,
thus $f(h_i)$ can be approximated by
the trailing edge of the activation curve:
$$d_B=pf(h_2)\approx p{\theta_2\over \theta_2 + h_2}
\ \ , \eqno (3.2)$$
where
$$h_2 = a_1 + (z-1)a_3\ \ . \eqno (3.3)$$
Level 3 is assumed to be in a virgin or near-virgin state.  Thus
$ a_3 \ll a_2$.
Since level 1 is activated and level 2 suppressed,
$a_2 < a_1$;
therefore, if $z$ is not too large,
$$h_{2}  \approx a_1\ \ . \eqno(3.4)$$

To find the approximate steady-state
values, we substitute Eq. (3.4) into (3.2).
Solving for $a_1$, we find
$$a_{1ss}\approx {\theta_2(p-d_B)\over d_B}\ \ .  \eqno (3.5)$$
Similarly, since level 1 is activated, $h_1\ll \theta_2$, and
$$d_B\approx pf(h_1)\approx p{h_1\over h_1+\theta_1}\ \ , \eqno (3.6)$$
where $h_1=za_2$.  Solving for $a_2$ yields
$$a_{2ss}\approx {d_B\theta_1\over z(p-d_B)} \ \ . \eqno (3.7)$$

Substituting $a_{1ss}$ and $a_{2ss}$ into the steady-state
conditions, i.e. Eqs. (2.3) with
$d{a}_1/dt = 0$ and $d{a}_2/dt=0$, yields
estimates for  $b_{1ss}$ and $b_{2ss}$:

$$b_{1ss}  \approx  {\theta_2p\over sd_B}
	\left[{(p-d_B)\over d_B} d_A + d_C\theta_1\right]\ \ .
\eqno (3.8a)
$$

$$b_{2ss} \approx {\theta_1\over sz}
\ \left[{p \over (p-d_B)} d_A+d_C\theta_2
{p \over d_B} \right]\
\ . \eqno (3.8b) $$

\skip
The steady-state values for the {\it antibody} populations,
Eqs. (3.5) and (3.7), turn out to be essentially
the same as the estimated {\it clone} sizes in the
corresponding
localized memory state in the B cell
Cayley tree model (Weisbuch et al., 1990).
This was to be expected, since it is the {\it field}
that determines clone size.
In the case of the B model, the field consists
of the B cell population levels; whereas, in the AB Tree model,
the field consists of the antibody populations.

\skip
For the standard parameter set, the approximate steady-state
values are $b_{1ss}=21,000$, $a_{1ss}=10,000$,
$b_{2ss}=2,000$, $a_{2ss}=10$.
The accuracy of these approximations was tested
by numerical calculation of
the {\it exact} steady-state values.
The approximate values were found to be
within 2\% of the numerical values.
Notice that in an immune state, the antibody population
of the memory level, $a_1$, is $z \theta_2 /\theta_1 $
times larger than the antibody, $a_2$,
of the sustaining, suppressed level.
The B cell populations, however, differ by a factor
approximately equal to $z$.
Thus, the clone size of the suppressed population
for $z=2$, for example, is only half the size
of the activated population.
In the case of the two-clone model ($z=1$),
$B_1$ is only slightly
larger than the $B_2$ in a stable immune state.

\skip
For the immune steady-state to remain localized,
level 3 must not become activated,
(i.e., $pf(h_3)<d_B$ or $h_3<\theta_1 {d_B\over{p-d_B}}$). Also,
level 4 is assumed to be near-virgin, so that
$$h_3=a_2+(z-1)a_4\approx a_2\ \ . \eqno (3.9)$$

Substituting the steady-state value for $a_2$
into Eq. (3.9)
yields as a necessary condition for this localized state
$$z>1 \  \ . \eqno(3.10)$$
Thus, for a localized memory to remain
localized, there must be
more clones in level 2 than in level 1.
This is in agreement with the results of Weisbuch et al. (1990).

\skip
We now estimate the steady-state
values for the level 3 populations.
Note that, since $b_3$ is assumed to be
near-virgin, $m$ is not negligible in this case.
Approximating $f(h_3) \approx f(a_2)$ by its rising part,
or
$$f(h_3)\approx {a_2\over a_2 + \theta_1}
\ \ . \eqno(3.11)$$
{}From Eq. (2.3), at steady-state
$$b_3=
	{	m\over d_B-p \left( {a_2\over a_2 + \theta_1} \right)  }\ \ . \eqno(3.12)
$$
Substituting $a_{2ss}$ yields
$$b_{3ss} \approx
	{m\over {d_B}}{z\over{(z-1)}}\
	\left[
		{ 1 + {d_B\over {z(p-d_B)}}}
	\right]
	\ \ . \eqno(3.13)$$
The corresponding steady-state antibody concentration is given by

$$
a_{3ss}=
	{  sf(h_3)b_{3ss}  \over
	(d_A+d_Cb_{3ss})  } \ \ . \eqno(3.14)$$
Substituting Eqs. (3.7), (3.13)
and (3.11) into (3.14) yields
$$a_{3ss} =
	{  sm(z/(z-1))  \over
	[d_Az(p-d_B)+d_Cd_B\theta_1]
	}
		\ \ . \eqno(3.15)$$

Notice that
$$\lim_{z  \to\infty} b_{3ss} = m/d_B\ \ ,
\eqno(3.16)
$$
and for large $z$, $b_3$ is nearly virgin
(${d_B\over{(p-d_B)}}=1$ for our standard parameters),
consistent with the condition for a localized state.
However, as $z$ increases,
$a_{2ss}$ decreases (see Eq. 3.7) and
$a_{4ss}$ increases
until the assumption of
Eq. (3.9) becomes invalid.
This will be shown explicitly in Section 6 using
numerical methods, and is illustrated in Fig.~4.
Again the accuracy of the approximate steady-state
populations
at level 3 were compared to their
numerically determined values.
For the standard parameter values, the estimated
values, $b_{3ss}$ and $a_{3ss}$, were 68\% and 62\%
of the numerical values.
As $z$ is increased from 10, the standard value,
this error increases significantly.
For example, at $z=15$, the
approximate steady-state populations
are only 34\% and 31\% of the numerical values.
Thus, the estimated steady-state values for level 3
are only valid for relatively small values of $z$.

Initially, we had calculated the steady-state
values for levels 1 and 2 assuming that
$$h_2=a_1+(z-1)a_3\approx a_1\ \ . $$
In order for this assumption to hold,
$a_{1ss}\gg(z-1)a_{3ss}$, or

$$\theta_2 \gg {mszd_B\over (p-d_B)p[d_Az(p-d_B) +d_Cd_B\theta_1]}\ \ . \eqno
(3.17)$$
For our standard parameters,
this condition is easily met.
Thus, in certain parameter regimes the
AB Tree model has localized steady-states.
The analysis thus far, however, has not put any conditions
on the stability of these states.
\skip
\leftline{\bigbf 4. STABILITY ANALYSIS}
\skip
We next find conditions under which the
localized immune steady-state is {\it stable}.
Stability analysis is greatly simplified
if we continue to use the approximations

(i)   \ \ $h_1\ll\theta_2$,\par
(ii)  \ \ $h_2\gg\theta_1$, and \par
(iii) \ \ $a_1\gg(z-1)a_3$.\par

\smallskip
Using (i) and (ii), we can approximate the
activation function for
levels 1 and 2 by the rising and
falling parts of $f(h)$, respectively.  Approximation (iii)
allows us to ignore population dynamics beyond level 2.
Thus, near the localized state, the model reduces to the
following four-dimensional form:

$$\eqalign{
{db_1\over dt} &= m + p\left({ za_2\over za_2+\theta_1}  \right) b_1 -
d_Bb_1\cr
\noalign{\vskip.1truein}
{da_1\over dt} &= s\left({ za_2\over za_2+\theta_1 } \right) b_1 - d_Aa_1 -
d_Ca_1za_2\cr
\noalign{\vskip.1truein}
{db_2\over dt} &= m + p\left( { \theta_2\over a_1+\theta_2} \right) b_2 -
d_Bb_2\cr
\noalign{\vskip.1truein}
{da_2\over dt} &= s\left( {\theta_2\over a_1+\theta_2} \right) b_2 - d_Aa_2 -
d_Ca_1a_2\cr
}
\eqno(4.1)
$$
Since it assumes no level 3 interactions, this model consists of
a single, first level clone and $z$ clones at level 2.
We shall refer to this reduced model as the ``star" model.

To linearize these equations
about the steady-state, we compute the Jacobian
$$J=\left[ \matrix{
	{\partial b_1\over \partial b_1}&\cdots&{\partial b_1\over \partial
a_2}\cr
	{.\atop .} & & {.\atop .}\cr
	{.\atop .} & & {.\atop .}\cr
	{\partial a_2\over \partial b_1}&\cdots&{\partial a_2\over \partial
a_2}\cr }
\right]$$
\vskip.15truein

$$=\left[\matrix{
	{pza_2\over za_2+\theta_1}-d_B & 0 & 0 &{pz\theta_1b_1\over (za_2 +
\theta_1)^2}\cr
\noalign{\vskip.1truein}
s {za_2\over za_2+\theta_1} & -d_A-d_Cza_2 & 0 & {sz\theta_1b_1\over \left(
za_2+\theta_1\right)^2}-d_Cza_1\cr
\noalign{\vskip.1truein}
0 & {  -p\theta_2b_2\over (a_1+\theta_2)^2  } & {p\theta_2\over a_1+\theta_2}
-d_B & 0\cr
\noalign{\vskip.1truein}
0 &  {  -s\theta_2b_2\over (a_1+\theta_2)^2  } -d_Ca_2 & {s\theta_2\over
a_1+\theta_2} & -d_A-d_Ca_1\cr}
\right]\ \  \eqno (4.2)$$
and evaluate it for the localized steady-state values
($a_{1ss}$, $a_{2ss}$, $b_{1ss}$, $b_{2ss}$)
given by Eqs (3.7) and (3.9)-(3.11).
Notice that when these substitutions are made,
two of the diagonal terms vanish:
$$J=\left[\matrix{
	0 & 0 & 0 &{pz\theta_1b_{1ss}\over (za_{2ss} + \theta_1)^2}\cr
\noalign{\vskip.1truein}
d_Bs \over p & -d_A-{{d_Cd_B\theta_1}\over {(p-d_B)}} & 0 &
{sz\theta_1b_{1ss}\over \left( za_{2ss}+\theta_1\right)^2}-d_Cza_{1ss}\cr
\noalign{\vskip.1truein}
0 & {  -p\theta_2b_{2ss}\over (a_{1ss}+\theta_2)^2  } & 0 & 0\cr
\noalign{\vskip.1truein}
0 &  {  -s\theta_2b_{2ss}\over (a_{1ss}+\theta_2)^2  } -d_Ca_{2ss} &
{d_Bs}\over p & -d_A-{d_C\theta_2(p-d_B)\over d_B}\cr}
\right]\ \  \eqno (4.3) $$

The eigenvalues, $\lambda$, of $J$ can be found by solving
the characteristic equation
$$p=det[\lambda I - J] =0\ ,  \eqno(4.4)$$
or,
$$p = c_0 + c_1 \lambda + c_2 \lambda^2 + c_3 \lambda^3 +
c_4 \lambda^4 = 0\ \ , \eqno (4.5)$$
where $p$ is the characteristic polynomial
and $c_i$'s are the coefficients of the characteristic equation.
The coefficients are as follows:

$$
\eqalignno{
c_0=&d_B^2(p-d_B)^2(-d_Ad_B+pd_A+d_Bd_C\theta_1)[d_Ad_B+(p-d_B)d_C\theta_2]&(4.6)\cr
\noalign{\vskip.1truein}
c_1=&p(p-d_B)\{-d_Ad_B^2d_C\theta_1(2d_B-p)+(p-d_B)^2[2d^2_Ad_B\cr
&-d_Ad_C\theta_2(2d_B-3p)+2d_Bd_C^2\theta_1\theta_2]\}&(4.7)\cr
\noalign{\vskip.1truein}
c_2=&
  d^2_Ad_B(p-d_B)(-2p^2+2pd_B-d^2_B) +d_Ad^2_Bd_C\theta_1(d^2_B-d_Bp+p^2)  \cr
& +d_Ad_C\theta_2(p-d_B)^2(3p^2-3d_Bp+d^2_B)
+d_Bd_C^2\theta_1\theta_2(p-d_B)^3 &(4.8)\cr
\noalign{\vskip.1truein}
c_3=&p^2[-2d_Ad_B(1-p)+d_B^2d_C\theta_1+d_C\theta_2(p-d_B)^2]&(4.9)\cr
\noalign{\vskip.1truein}
c_4=& p^2d_B(p-d_B)\ \ . &(4.10)\cr
}
$$

\skip
Surprisingly, the characteristic equation is
{\it independent of the coordination number, z}.
(All $z$ terms of the characteristic polynomial
are common factors.)
This is still true when the {\it exact} Jacobian is used
(that is, the Jacobian taken using the full $f(h)$ and not
simply it's rising and falling parts).
What this implies is
that stability is insensitive to the
asymmetry in the model due to the fact that
there is one level 1 clone and $z$ level 2 clones.
Thus, if a localized memory state becomes unstable as $z$
is changed, it is due to the interactions with
level 3 populations.
This is explored further in Section 6.
\skip
If we set all parameters to constants and choose one
parameter as a variable,
the characteristic equation
allows us to predict stability as a function of that variable.
For example, if we vary the antibody death rate, $d_A$,
leaving all other parameters at their standard values,
we get the characteristic equation as a function
of $d_A$ with the coefficients
$$ c_0 = 6.25 + 6.31 d_A + 0.0625 d_A^2\ \ , \eqno (4.11)$$
$$ c_1 = 25 + 50 d_A + 0.25 d_A^2\ \ , \eqno (4.12)$$
$$ c_2 = 25 + 175.75 d_A + 1.25 d_A^2\ \ , \eqno (4.13)$$
$$ c_3 = 101 + 2 d_A\ \ . \eqno (4.14)$$

To find stability conditions in this case,
it is not strictly necessary
to find the eigenvalues;
inspection of the coefficients of the characteristic
polynomial is sufficient.  The characteristic polynomial
is stable if the following conditions are satisfied
(Li\'enard-Chipart Theorem (Fortmann and Hitz, 1977)):
$$ c_i > 0 \ \ \forall \ \ i \ \ , \eqno (4.15)$$
and
$$ c_3 c_2 c_1 > c_3^2 c_0 + c_1^2 \ \ .  \eqno (4.16)$$
Conditions (4.15) are always met in this example since $d_A > 0$.
Condition (4.16)
predicts that the localized
state is stable for values of
$$d_A>0.0025 \ \ . \eqno (4.17)$$
If we repeat this analysis using the exact Jacobian,
we find the slightly stronger condition
$$d_A>0.0047 \ \ . \eqno (4.18)$$
Our estimated value of $d_A$ is 0.05, and thus with
our standard parameters, the localized immune state
is stable. Note, however, that
if the antibody lifetime is too short,
network interactions leading to localized memories
cannot be sustained.
\skip
Similarly, conditions can be derived by varying other
parameters. Setting $d_A = 0.05$ and freeing $d_C$ yields
$$ c_0 = .00015625 + 31.5625 d_C + 62500 d_C^2 \eqno (4.19)$$
$$ c_1 = .000625 + 250000 d_C + 250 d_C^2 \eqno (4.20)$$
$$ c_2 = .003125 + 878.75 d_C + 250000 d_C^2 \eqno (4.21)$$
$$ c_3 = 10100 + 0.1 d_C \eqno (4.22)$$
or
$$ d_C > 0.0060 \ \ . \eqno (4.23)$$
This condition remains essentially
unchanged when the exact Jacobian is used.
Again, with the estimated value, $d_C=0.01$, our analysis
predicts a stable localized immune state.
\skip
\leftline{\bigbf 5. LOCALIZED STATES AT OTHER LEVELS}
\skip
As previously noted, a localized state with level 1 high
is a special case
in that clones at lower levels need not be considered.
Localized states at other levels are of interest as a
generalization of the previous
analysis as well as their potential biological relevance.
For example, a localized state with level 2 high and
level 1 low or intermediate has been referred
to as a ``tolerance attractor"
(Weisbuch et al., 1990; Neumann and Weisbuch 1992a).
A high level of $Ab_2$
suppresses the primary antibody response rendering the
network unresponsive, or ``tolerant", to antigenic challenge.
\skip
To find the conditions for a localized state with a level other
than level 1 high requires a similar analysis.
Consider a state with $a_i$ high (e.g. $a_i \approx \theta_2$).
Level $i-1$, experiencing a field of $a_{i-2} + (z-1)a_{i}$, will
be far into the suppressive range of the dose-response curve;
consequently, $a_{i-1}$ will be very low, perhaps near-virgin.
Thus, the only assumption in section 3 that changes is the
the approximation for the level {\it i} field

$$h_i = a_{i-1} + (z-1)a_{i+1}\approx (z-1)a_{i+1}\ \ . \eqno(5.1)$$

The steady-state values are then given by

$$a_{iss}\approx {\theta_2(p-d_B)\over d_B}\   \eqno (5.2)$$

$$a_{(i+1)ss}\approx {d_B\theta_1\over (z-1)(p-d_B)} \  \eqno (5.3)$$

$$b_{iss}  \approx  {\theta_2p\over sd_B}
	\left[{(p-d_B)\over d_B} d_A + d_C\theta_1\right]\
\eqno (5.4)
$$

$$b_{(i+1)ss} \approx {\theta_1p\over s(z-1)(p-d_B)}
\ \left[d_A+d_C\theta_2 {(p-d_B)\over d_B} \right]\ \ . \eqno (5.5)
$$
An example of a ``tolerance" attractor (a localized steady-state
at level 2, with a sustaining population at
level 3) is shown in Fig. 2.

\skip
The necessary condition corresponding to Eq. (3.10)
for this localized state is
$$z>2 \ . \eqno (5.6)$$
The unit increase in the condition on z is a direct consequence
of network structure.
A similar condition has been found for more general structures
(Neumann and Weisbuch 1992b).
This condition is really the same as Eq. (3.14); that is,
there must be more than one connected clone descending down the
Cayley tree. Thus, the simplest structure
which can support tolerance is a tree with coordination
number $z=3$ (see Fig. 1).

\skip
\leftline{\bigbf 6. NUMERICAL BIFURCATION ANALYSIS}
\skip
Having established some approximate conditions for
stable, localized
steady-state network behavior from the star model,
we wish to know what happens to these states as
model parameters are changed and
the stability conditions are violated.
In this section, we
analyze steady-state behavior of a more complete
model of a Cayley tree
model using a numerical
bifurcation analysis software package, AUTO
(Doedel, 1981; Taylor and Kevrekidis, 1990).
The following numerical work was performed on a
ten-level Cayley tree model (fields due to levels 11 and
beyond are assumed to be zero).
The result is a 20-dimensional system of equations
(one equation for each B cell and antibody population
at each level).
In analyzing bifurcations that occur as $z$ is varied,
we treat $z$
as a continuous variable.
However, strictly speaking, Cayley trees are only defined
for integer values of $z$.

\skip
\leftline{\bigbf 6.1 Nondimensional Model}
\skip
First, we nondimensionalize the model equations to reduce the
number of model parameters.
For comparison, we have attempted
to choose dimensionless units which
are roughly equivalent to those in De Boer and Perelson's (1993a,b)
analysis of two-clone models.
Accordingly, the time scale is based upon the B cell
lifetime, i.e. $T=td_B$.
We scale the antibody concentration by a factor,
$\alpha=\sqrt{\theta_1 \theta_2}$, which
corresponds to the concentration
of antibody which leads to maximum crosslinking (activation).
We then scale the B cell
population by a factor, $\beta = (d_A \alpha)/s$,
the concentration of B cells required to sustain
a steady-state population of $\alpha$ antibodies
(at maximum activation and ignoring
complex formation).
The remaining quantities, $h_i, \theta_1,$ and $\theta_2$
are scaled by $\alpha$.
The nondimensional dynamical equations become
$$ H_i = \sum_j J_{ij}A_j \ \ ,\eqno(6.1) $$
$$
f(H_i) =
	{H_i\over(\Theta_1+H_i)}
	 \
	{\Theta_2\over (\Theta_2+H_i)}\ \ ,\eqno(6.2) $$
where $\Theta_1 = \theta_1/\alpha$, and
$\Theta_2 = \theta_2/\alpha$.
$$
\eqalign{
{dB_i\over dT} &= \sigma + (\rho f(H_i)-1) B_i\ \ ,\cr
\noalign{\vskip.1truein}
{dA_i\over dT} &= \nu f(H_i)B_i -(\delta +\mu H_i) A_i\ \ ,\cr
}
\eqno(6.3) $$
where
$$A_i=a_i/\alpha,
\quad B_i=b_i/\beta,
\quad \delta=d_A/d_B,
\quad \sigma=m/(\beta d_B),
\quad \rho=p/d_B, $$
$$ \nu={\beta s}/(\alpha d_B),
\quad \mu = (\alpha d_C)/d_B,
\quad \alpha= \sqrt{\theta_1 \theta_2},
\quad \beta= (\alpha d_A)/s. $$
The corresponding standard, non-dimensional parameter values
are $\delta=0.1$, $\sigma=0.04$,
$\rho=2$, $ \nu=0.1$,
$\mu = 20$, $\alpha= 1000$, and
$\beta= 50$.
\skip
\leftline{\bigbf 6.2 Connectivity Dependence of Localized Steady States}
\skip
The connectivity parameter $z$ is the only new feature
added to the basic two-clone AB model (De Boer et al., 1993a,b).
We introduce a network structure to the two-clone model
when $z \geq 2$.
Thus, we first investigate the dependence of
the localized steady-state on the connectivity parameter, $z$.
Figure 3 shows
the $z$-dependence of two different
localized steady-states, one with level 1 high (a localized
immune state) and one with level 2 high (a
localized tolerance attractor).
With all other model parameters set to the standard values,
conditions for stability of
these states are $1\leq z \leq 15$ and
$2\leq z \leq 16$, respectively.
For $z=1$, these steady-states correspond to the ``HM"
and ``MH" states
in the two-clone AB model (De Boer et al., 1993a,b).
As discussed in Sections 3 and 4,
the upper limits on $z$ are imposed from interactions
with level 3 populations.
As $z$ gets large, the approximation for the field
at level 3, (
$h_3=a_2+(z-1)a_4\approx a_2$,
Eq. (3.9)), breaks down.
As the field at level 3 increases,
it's clonal population increases until
it begins to stimulate higher-level clones.
\skip
Both of the localized states
(with levels 1 and 2 non-virgin)
exist as ``isolated" solutions;
that is, as $z$ is varied
the steady-states do not branch into other
attractors, but rather loop back on themselves.
In Fig. 3, the lower branches, indicated by the dashed lines,
are unstable.

\skip
\leftline{\bigbf 6.3 Extended Localization and Percolation Attractors}
\skip
The stability of the localized immune
state is independent of $z$ in the two-level, star model;
therefore, it is the interactions with deeper
levels in the immune network which destroys the
immune state.
As $z$ is increased, clone 3 begins to expand far enough
above virgin levels to stimulate proliferation of level 4 clones.
We refer to the loss of localization as {\it structural},
since system steady states are dynamically stable,
while localized states are lost due to changes
in the model structure (connectivity).
As discussed above,
the assumption that
$h_3=a_2+(z-1)a_4\approx a_2$, Eq. (3.9),
only holds for small $z$.
In Fig. 4, the two components of the field
experienced by level 3 clones is plotted against $z$.
As $(z-1) A_4$ becomes comparable $A_2$, the
localized immune state is lost.
\skip
For high $z$, stable steady-states still exist,
but these states correspond to ``extended localization"
(Neumann and Weisbuch 1992a)
and ``percolation" attractors, where
many levels are maintained at high populations.
Figure 5 shows the dynamical trajectory which results
when the localized memory state is lost ($z=16$).
The initial system state was chosen to be
the immune state for $z=15$.
At t=0, $z$ was increased to $z=16$, and Eqs. (6.3)
were integrated numerically.
Since no localized immune steady-state now exists,
the trajectory moves into a new basin of attraction
(in this case an extended localization with B cells at levels
1 and 4 high, 3 and 5 intermediate, and deeper levels near-virgin).
This state exists for a slightly larger range,
$z\leq 19$ (Fig. 3).
If we continue to increase $z$,
activation cascades further down the network resulting
in a percolation attractor.

Percolation attractors can {\it coexist} with
localized steady-states in the AB Tree model.
Notice in Fig. 3
that for $2<z<16$, the extended localized attractor coexists
with a tolerance attractor.
Thus,
when the localized memory state disappears as $z$ is increased,
it does not spawn a new
attractor; trajectories simply approach other existing
(immune, virgin, or one of the percolation) attractors -
depending upon initial conditions.

\skip
\hang{\bigbf 6.4 Dependence of the Localized Steady State
on Antibody Dynamics}
\skip
The inclusion of antibody populations as state
variables in the Cayley tree model introduces two important
parameters,
the antibody death rate, $d_A$, and
the complex removal rate, $d_C$, i.e., dimensionless parameters
$\delta$ and $\mu$, respectively.
Varying these parameters can
change steady-state behavior into chaotic behavior.
The loss of localization in this case is {\it dynamical},
since nearly all stable steady state behaviors
(including percolation attractors) are lost
with changes in the dynamical variables.

\skip
The stability of the localized immune steady-state
as a function of $\delta$
(the ratio of antibody/B cell death rates)
is shown in Fig. 6.
If $\delta$ is increased from its standard value of
0.1, the eigenvalues become
increasingly negative, i.e. more stable (not shown).
As $\delta$ is decreased from 0.1,
a Hopf bifurcation occurs
at $\delta \approx 0.0136$,
and the steady-state goes unstable.
For the ``standard" parameter value $d_B=0.5$,
this corresponds to the condition
$d_A = 0.0068$,
which is close to the estimate
provided from linear stability analysis of
the star model (Eq. (4.18)
$d_A > 0.0047$).
The Hopf bifurcation branch
consists of unstable limit cycles,
while continuation of the primary branch follows
an unstable steady-state.
Most attractors in this region are chaotic,
although some are steady-states with levels other
than level 1 high.

\skip
Localized states can also loose stability if the
complex formation parameter, $\mu$,
becomes too small.
Figure 7 is a bifurcation diagram of
the level 1 (immune) localized steady-state
with $\mu$ as the bifurcation parameter.
Beginning with the standard value ($\mu = 20$),
the steady-state
becomes unstable at the Hopf bifurcation
as $\mu$ drops below 12.45.
For the standard parameter set,
instability corresponds to the condition $d_C < 0.00623$.
This, also, is in close agreement to the estimate of 0.0060
from linear stability analysis of the star model (Eq. (4.23)).
Past the Hopf bifurcation, the steady-state is unstable
with 2 complex eigenvalues, both having positive real parts.
At the saddle-node bifurcation,
an additional positive, real eigenvalue
appears; thus, along the lower branch of the bifurcation curve
the system has 3 eigenvalues with positive real part.
At $\mu=12.1$, the complex eigenvalues re-cross
the imaginary axis, but the single unstable eigenvalue
persists becoming increasingly
unstable with increasing $\mu$.
The branches from the two Hopf bifurcations
consist of unstable limit cycles.
In the region past the first Hopf bifurcation, i.e. $\mu <12.45$,
system attractors, other than the virgin state, appear
to be chaotic. For $\mu > 12.45$, the immune state is stable and
surrounded by an unstable limit cycle. This unstable limit cycle,
along with its stable manifold, define the basin of
attraction of the stable immune state.
\skip
Figure 8 is a two-parameter continuation of the localized
immune state.
Assuming standard values for the other parameters,
this diagram shows the combinations of
$\mu$ and $\delta$ for which the localized immune
state at level 1 exists, as well as whether it is stable.
No localized immune steady-state
exists below the saddle-node curve.
The two broken lines indicate the boundaries
for Hopf bifurcation curves (HB-1 and HB-2).
The steady-state is stable only above the first Hopf
bifurcation (HB-1).
In the region between the saddle-node and
Hopf bifurcation curves, only unstable
steady-states exist.
The one parameter continuations in Figs. 6 and 7
project onto this diagram as
a vertical line at $\mu=20$
and a horizontal line at $\delta=0.1$, respectively.
This diagram qualitatively corresponds to Fig. 3 in
Perelson and Weisbuch (1992) and De Boer et al. (1993a,b).
\skip
Figure 9 is a two-parameter continuation of the
saddle-node and Hopf bifurcations of the
localized steady-state at level 1
varying one dynamical parameter, $\mu$, and
the connectivity parameter, $z$.
As $\mu$ is lowered, the localized steady-state
exists and is stable
for a decreasing range of $z$.
At approximately $\mu=12.5$, the steady-state
becomes unstable
for all values of $z$,
and remains unstable for all $\mu<12.5$.
The loss of stability occurs via a Hopf
bifurcation.
The limit cycles that appear are unstable.
Again, network connectivity, $z$, mostly determines
the existence of the localized state and
the dynamical parameter
primarily determines the stability of the steady-state.
The one parameter continuations in Figs. 3 and 7
project onto this diagram as
a vertical line at $z=20$
and a horizontal line at $\mu=20$, respectively.
\skip
\leftline{\bigbf 6.5 Chaotic Attractors}
\skip
As system parameters are varied past the Hopf
bifurcations, the dynamics can become chaotic.
To study the dynamics we use the dimensional
equations (2.3).
In Figs. 10a-h, a series of time plots and
phase portraits are
shown for $z=10$ as $d_C$ is decreased past the critical value
of 0.00623, and the localized steady-state becomes unstable.
Beginning with the standard parameter set
($d_C=0.01$),
the localized memory state is asymptotically stable
(Figs. 10a,b).
As $d_C$ is lowered toward the Hopf bifurcation at
$d_C \approx 0.00623$,
the basin of attraction for the localized steady-state
shrinks.  Because the steady-state is surrounded by
an unstable limit cycle, a large enough perturbation
will cause trajectories to move away from the steady-state
and approach another attractor.
This is illustrated in Figs. 10c and d for $d_C=0.0067$.
Here a large perturbation was given and the trajectory
slowly moves away from the steady-state, goes through a
transient, and then
approaches an apparently chaotic attractor.
This attractor resembles the Lorenz attractor
(Sparrow, 1982)
in that the trajectory spirals around two stable states, one
with  $a_1$ high, the other with $a_2$ high.
Just past the Hopf bifurcation ($d_C=0.0060$),
chaotic trajectories are also observed.
In Figs. 10e and f it is seen that
the trajectory often returns to the
region in state space near the unstable steady-state.
As $d_C$ is reduced further,
the attractor becomes increasingly dispersed
in state space.
This is illustrated in Figs. 10g and h
for $d_C = 0.001$.

\skip
The time course of a typical chaotic
attractor is shown in Fig.~11.
At time t=100, a large dose ($10^5$) of $Ab_1$
is ``injected" into the system.
As can be seen, this causes $Ab_1$ and $Ab_2$
to stop fluctuating for about 50 days, while leaving $Ab_3$,
$Ab_4$ and $Ab_5$ fluctuating (Fig. 11a).
After 50 days, the system relaxes back into the fully chaotic
state. Doses of injected antibody of order $10^4$ or less
have little effect on network dynamics and the fluctuations
continue unabetted.
For larger $z$, a larger dose is needed to disturb network
dynamics due to the large number of connected clones
at level 2 (Fig. 11b).

\skip
\leftline{\bigbf 6.6 Localized Chaos and Limit Cycles}
\skip
In the two-clone AB model, stable limit cycle attractors
were found
over a wide parameter range
(De Boer et al., 1993a,b).
With large amplitude oscillations
in level 1 and 2 antibody populations,
however, level 3 would be expected to be
stimulated past the virgin
threshold, leading to percolation or chaos.
Indeed, when the system parameters were set to
those of the oscillatory regime of the two-clone model
(e.g. $\mu<12.68$, dimensional value $d_C<.00634$),
chaotic dynamics spread throughout the network, even for z=2.
Moreover, even in the parameter regime in which the
two-clone AB model exhibits limit cycle behavior ($\mu<0.18$),
the AB Tree model shows chaotic behavior for small $z$.
Thus, the introduction of even the most minimal
network structure, a linear chain, disrupted
the limit cycle oscillations observed in the two-clone model.
\skip
The potential exists for localized oscillatory states
if the oscillations of $a_2$ remain sufficiently small,
i.e. below the threshold for activating level 3,
such that the condition $a_{2max} < {d_B \over {(p-d_B)}} \theta_1$
is fulfilled.
On the other hand,
the oscillations of $a_2$ must be smaller then the suppressive threshold
in order to sustain oscillations at level 1,
i.e. $h_{1max} = z a_{2max} = {{(p-d_B)} \over d_B} \theta_2$.
By combining the above two expressions we obtain a sufficient condition
for localized oscillations,
$$ z \geq  {\theta_2 \over \theta_1} \ \ , \eqno (6.4) $$
enough so that $h_1 = z a_{2} \le {{(p-d_B)} \over d_B} \theta_2$.
For our standard parameter set, this condition is
satisfied for $z>100$.
\skip
Indeed, stable limit cycles have been found in high connectivity
parameter regimes.
One such attractor is shown in Fig. 12a,b.
These limit cycles are only structurally
stable when the bone marrow
source term is extremely small so that virgin B cell
clones are too small to be activated easily ($m = .000025$).
As $d_C$ is increased, the limit cycle becomes unstable,
and system dynamics are characterized by long-lived
oscillatory transients
which do not activate higher level clones (Fig.12c,d).
At even higher values of $d_C$,
a localized chaotic attractor appears (Fig. 12e,f),
where chaotic oscillations at levels 1 and 2 do {\it not}
substantially disturb the near-virgin populations
deeper in the network.

\skip
\leftline{\bigbf 7. DISCUSSION}
\skip

\skip
\leftline{\bigbf 7.1 More Complex Network Structures}
\skip

We have studied the behavior of antibody--B cell immune
networks that have the topology of a Cayley tree.
A Cayley tree is a
homogeneous network, without loops, in which every node
is connected to precisely $z$ others. The Cayley tree
is clearly only an approximation to real immune network
topology. While each clone in
a network may be connected to $z$ others (on average)
it is unlikely that all clones would ever be connected to
exactly $z$ others. Natural
IgM antibodies in neonatal mice, when tested in binding
assays,  exhibit highly variable
reactivities (Holmberg et al., 1984; Holmberg, 1987).
Many of the antibodies are found to be highly multireactive,
while others are specific.  Thus, at least in this example
a homogeneous topology does not seem to exist.
The effect of variable
connectivities on system attractors has
been studied for the B cell Cayley tree model
(Neumann and Weisbuch 1992b) but not
on the AB Tree model.
\skip
Further evidence of network structures that differ
from the Cayley tree model comes from functional distinctions
between classes of second level antibodies
(Jerne, 1974; Jerne et al., 1982).
Primary antibodies ($Ab_1$'s) recognize epitopes
of an antigen.
Secondary antibodies ($Ab_2$'s), can  recognize either
idiotopes or paratopes of $Ab_1$.
If an $Ab_2$ recognizes an idiotope outside the binding
site it is
classified as an $Ab_{2\alpha}$ antibody, while if it
recognizes the paratope of $Ab_1$ it is referred to as
an $Ab_{2 \beta}$ or ``internal image" antibody,
since it mimics the shape of the
original antigenic determinant.
Internal images are not accounted for in
a Cayley tree structure since, as we show below, they generate loops.
\skip
An internal image could be added to a network model.
Consider network with coordination number $z$ and
an external antigen as it's root (see Fig.~13).
An internal image
would be indistinguishable from the antigen itself.
If we allow a fraction
$\mu$ of the second level antibodies to be internal
images $Ab_{2\beta}$ of the antigen,
the fields become:
$$h_1 = Ag + z[\mu a_{2\beta} +  (1-\mu) a_{2\alpha}] \ \ , \eqno (7.1)$$
where Ag is the effective antigen concentration.
If we assume that the $Ab_1$'s represent the dominant
idiotypic interactions for an internal image,
the field for the internal images is
$$h_{2\beta} = za_1 \ \ , \eqno (7.2)$$
(If one were to include further connectivity,
a separate population of $Ab_{3\alpha}$'s would need to be added.)
The $Ab_{2\alpha}$'s would retain a tree-like connectivity:
$$h_{2\alpha} =  a_1 + (z-1)a_3\ \ . \eqno (7.3)$$
The inclusion of internal images
violates the tree structure, and the dynamics of
the $Ab_{2\alpha}$'s and $Ab_{2\beta}$'s must now be treated separately.
\skip

K\"ohler subdivides $Ab_{2}$ antibodies differently
than Jerne et al. (1984)  by
defining a ``network antigen" as
an  $Ab_{2}$ that can be used for vaccination
(K\"ohler et al., 1989; K\"ohler, 1991).
Network antigens do not necessarily meet the immunochemical
criteria of internal images, but still are capable of
inducing biologically beneficial immune responses.
Network antigens and internal images have been
used to prime an immune response
without exposing an animal to the antigen itself
(K\"ohler et al., 1986;
Huang et al., 1988;
Raychaudhuri et al., 1990;
Bhattacharya-Chatterjee et al., 1990) and hence
have obvious use as potential vaccines.
Antibodies connected
in loops may be used to model the connectivity of
a network antigen (Fig.~13).
\skip
It is not only internal images and network antigens that
generate loops, but
as pointed out by Neumann and Weisbuch (1992b),
any recognition scheme based on complementary
shapes implies a network with loops.
For example, if $Ab_2$ and $Ab_4$ resemble each other,
they may both interact with $Ab_3$ and $Ab_1$ forming a four-membered
loop.
In the case of B models, Neumann and Weisbuch (1992b) have
used the window automata approximation (Neumann and Weisbuch,
1992a) to analyze the effects of simple loops on the existence and stability
of localized states.  Similar analyses remain to be done for
AB models.
\skip

\leftline{\bigbf 7.2 Oscillations  and Immune Memory}
\skip
Immune networks may be able to store
memories in the form of dynamic steady-states
(Farmer et al., 1986; Weisbuch, 1990; Weisbuch et al., 1990;
Behn et al., 1992). Generally, when networks are used to explain
memory to previous antigenic challenge the following
implicit hypotheses are made (cf., Weisbuch et al., 1990):

(i) The immune system is antigen-driven; that is,
prior to antigenic challenge, clones are in a
stationary, virgin state.

(ii) Antigenic challenge can force clones from the virgin
state into other states, such as those that correspond
to immune and tolerant attractors.

(iii) If the new attractors that the system is driven to
remain localized,
the network will be capable of storing memories of many
different antigens.
\skip
Some recent experimental data, however,
do not support the hypothesis that the immune
system is antigen-driven and that immunological
memory is stored in stable, localized steady-states.
\skip
Measurements of
naturally occurring antibody (NAb) concentrations {\it in
vivo} at various times show complex dynamics.
In the absence of external antigenic stimulation
individual NAb concentrations fluctuate irregularly
over time
(Lundkvist et al., 1989; Varela et al., 1991).
Based on Fourier spectra of rather limited time series,
Lundkvist et al. argue that the fluctuations appear to be
chaotic.
However, because the data are so limited it is
uncertain whether these fluctuations indicate the existence
of a chaotic attractor, a high-dimensional limit cycle
or are simply
the result of noise and perturbations about a
non-virgin steady-state.
In germ-free mice the number of activated lymphocytes
in the spleen and the serum level of IgM
are similar to the values measured in conventionally
raised animals (Hooykaas et al., 1984; Pereira et al., 1986).
These data as well as the Lundkvist data indicate that
the immune system is not in a rest
state in the absence of external antigen.

\skip
Coutinho (1989) has
argued that about 10-20\% of the immune system is organized into
an idiotypic network, or ``central immune system"
 that is active in the absence of external
antigen, and that the remaining 80-90\% of clones are
outside the network and constitute a ``peripheral immune system"
that is  responsible for immune responses to
foreign antigens. Thus, according to Coutinho, secondary
responses and hence memory would be non-network derived.
Whether a system in which clones and their anti-idiotypic clones
were in localized states and relatively unresponsive to
other activities in the network would correspond to the network
or non-network parts of the system is unclear. Clones in the
immune state could participate in rapid responses to antigen
characteristic of secondary immune responses.  However, while
in the immune state they would be activated and subject to
network interactions with their anti-idiotypic clone.
\skip
Lundkvist et al. (1989) did one additional experiment suggesting
that the fluctuations in NAb populations are not due to noise.
They  showed that the fluctuations in the serum concentrations of
natural antibodies with complementary idiotypes, which for
 notational simplicity, we call $Ab_1$ and $Ab_2$, could
be eliminated for three months by
injection of monoclonal antibodies with the
idiotypes carried by either $Ab_1$ or $Ab_2$.
Interestingly,
the dynamics of serum antibodies with
{\it unrelated} idiotypes remained relatively undisturbed
and continued to fluctuate
(Lundkvist et al., 1989).
This might suggest that dynamical network activity remains
localized in the immune network
since dynamical behavior in only part of
the immune system was noticeably changed.

\skip
We performed a similar experiment of injecting $Ab_1$
in our Cayley tree model when in
a ``chaotic" parameter regime. We found, as did
De Boer et al. (1990, 1993b) for the two-clone AB model,
 that injection of high doses
of $Ab_1$ could eliminate oscillations in $Ab_1$ and
$Ab_2$ for a period of months (Fig.~11).  However, low or moderate
dose injections frequently would not lead to a loss of
oscillations, the outcome depending on parameters values
and the concentrations of antibodies present in the
system at the time of the injection.  Interesting,
however, is that when oscillations at the $Ab_1$ and $Ab_2$
levels were eliminated, the higher levels $Ab_3,\ Ab_4,$
and $Ab_5$ still oscillated
(see Fig.~11).
Thus in the AB Cayley tree model
we can reproduce this second feature of the Lundkvist
experiments that was not apparent in the previous
two-clone AB  models. Further, our model indicates
that the continued fluctuations in higher levels of the
tree, while fluctuations at levels 1 and 2 are eliminated,
do not indicate that network activity is localized. In
fact, this effect is seen in the model in the
chaotic/percolating parameter regime. The explanation
of this phenomenon in our model is that with  high $Ab_1$,
B cells at level 2 are suppressed and $Ab_2$ concentration levels
are brought very low by a combination of complex
formation with $Ab_1$ and lack of production by suppressed
B cells.  Since level 3 is influenced by both level 2 and
level 4,  it can continue to oscillate with $Ab_2$ very
low as long as level 4 can stimulate it. In the chaotic
regime, $Ab_4$ gets high enough to trigger level 3 and
continue the percolation to higher levels.

\skip
The  Lundkvist data suggests that if immune memory is
stored in dynamical attractors
they must be more complex than simple point attractors.
It is difficult to envision memory storage in
the global percolation and chaotic attractors
found in the AB Tree model;
however, the localized chaotic and limit cycle
attractors found
in section 6.6 could serve a localized memory role.
Although these attractors were only
found in very extreme parameter regions,
in many other parameter regimes,
transient oscillations around a steady-state may persist
for as long as the lifetime of a mouse. For example,
in Fig.~10c a large
perturbation around a stable immune state produces
slowly growing oscillations that last about 700 days.

\skip
Although the natural
state of an immune network might be oscillatory,
one would expect that if antigen drives the network
then the {\it time-averaged} $Ab_1$ population level
would be much higher after antigenic challenge than before
challenge.
Indeed, the immune response to some antigens
is oscillatory (Weigle, 1975; Romball and Weigle, 1982;
Hiernaux et al., 1982)
with the time-averaged antibody concentration remaining high
for many weeks or months after antigenic challenge.
The oscillations  are usually damped and may reflect a
slow return to a steady-state.
\skip
Thus, even if the immune system operates in an oscillatory
or percolation regime it is still possible for memory to be
stored dynamically.  If responses stay localized it is easy to
envision how both memory storage and memory recall would work.
If responses do not stay localized it is much more difficult to
see how the immune system could utilize dynamic memory.  But
this is not to say that it would be impossible.  Neural networks
of the Hopfield type store memory in a non-local manner and this
provides certain advantages if damage occurs to particular
parts of the network.
\skip
\leftline{\bigbf 7.3 Conclusions}
\skip
The AB Tree model differs from previous models
in that it adds a simple network structure
to the two-clone AB models and antibody dynamics
to B cell Cayley tree models.
We have shown that
the inclusion of antibody dynamics does not
change the general conclusion of Weisbuch et al. (1990)
that there can exist stable
localized memory states in a Cayley tree
immune network model.
\skip
Besides the immune, tolerant and extended
localized steady-states,
we have identified two other classes of localized
system attractors:
limit cycles and localized chaotic attractors.
Global system attractors
include virgin, percolation and chaotic attractors.
Percolation attractors
are stable steady-states where many, if not all, network
levels are non-virgin.
In the AB Tree model,
percolation attractors coexist with localized
memories in many parameter regimes.
\skip
The primary new variable introduced in the AB Tree model
from the two-clone AB models is the network
connectivity, or more precisely, $z$, the coordination
number of the tree.
As $z$ is increased,
stable localized steady-states disappear, and only
percolation and chaotic attractors remain (Fig.~3).
This breakdown of localization is due to
interactions with an increasing number of connected clones
at higher levels in the tree.
Chaotic attractors do not exist in the B cell
Cayley tree model.
In parameter regimes where the two-clone AB model
shows limit cycle behavior, the AB Tree model
exhibits chaotic behavior.
But, in highly connected networks, limit cycle
behavior reappears, along with an interesting
new type of system attractor - a localized chaotic attractor.
\skip
In the dynamical simulations presented here,
chaotic or oscillatory behavior usually
percolates indefinitely through all levels.
Information could not easily be stored in such
attractors.
However, based on the Lundkvist experiments
we believe it likely that oscillatory
or chaotic attractors exist in real immune networks (Section 7.2).
The AB Cayley tree
model leaves out important idiotypic interactions,
such as internal images,  and features such as
gearing-up (Segel and Perelson, 1989) and
separate spleen and blood compartments (Perelson and Weisbuch,
1992; De Boer et al., 1993a,b).
Whether including additional features in the model will
serve to localize the dynamics in the network remains to
be explored.

\skip
\leftline{\bigbf Acknowledgements}
\skip
This work was performed under the auspices of the U.S.
Department of Energy and supported by NIH grant
RR06555 (ASP),
Sullivan Fellowship (AUN),
Fogarty International Research Fellowship  NIH--F05-TW04809 (AUN)
and the Los Alamos Center for Nonlinear Studies (RWA).
We thank Rob J. De Boer
and Mark A. Taylor for many useful
discussions and criticisms.

\vfill\eject
\skip
\centerline{\bf LITERATURE}
\bigskip
{
\frenchspacing
\parskip=8pt
\def\ti#1{#1.} 
\def\bt#1{{\sl #1}} 
\def\vo#1{{\bf #1},} 
\def\pp#1#2{#1--#2} 
\def\bp#1#2{pp. #1---#2} 
\def\jo#1{{\sl #1},} 
\def\in#1{In: {\sl #1}} 

Behn, U., J. L. van Hemmen and B. Sulzer. 1992.
\ti{Memory in the immune system: Network theory including
memory B cells} (to appear).

Bhattacharya-Chatterjee, M., S. Mukerjee, W. Biddle,
K.A. Foon and Heinz K\"ohler. 1990.
\ti{Murine monoclonal anti-idiotype antibody as a potential
network antigen for human carcinoembryonic antigen}
\jo{J. Immunol.} \vo{145} \pp{2758}{2765}

Bona, C. and H. K\"ohler. 1984.
\ti{Anti-idiotypic antibodies and internal images}
\in{Receptor Biochemistry and Methodology}, C. Ventor, ed.,
\vo{14} 141--149.
Alan R. Liss Inc.: New York.

Celada, F. 1971.
\ti{The cellular basis of immunological memory}
\jo{Prog. Allergy} \vo{15} 223--267.

Celada, F. and P. E. Seiden. 1992.
\ti{A computer model of cellular interactions in the immune system}
\jo{Immunol. Today} \vo{13} \pp{56}{62}

Coutinho, A. 1989.
\ti{Beyond clonal selection and network}
\jo{Immunol. Revs.} \vo{110} \pp{63}{87}.

De Boer, R. J. 1983.
\ti{GRIND: great integrator differential equations}
University of Utrecht, The Netherlands.

De Boer, R. J. 1988.
\ti{Symmetric idiotypic networks: connectance and switching,
stability, and suppression}
\in{Theoretical Immunology, Part Two, SFI Studies in the
Sciences of Complexity},
A. S. Perelson, ed.,
Addison-Wesley: Redwood City, CA, pp. 265--289.

De Boer, R. J., I. G. Kevrekidis and A. S. Perelson. 1990.
\ti{A simple idiotypic network model with complex dynamics}
\jo{Chem. Eng. Sci.} \vo{45} \pp{2375}{2382}.

De Boer, R. J. and A. S. Perelson. 1991.
\ti{Size and connectivity as emergent properties of a developing
immune network} \jo{J. Theor. Biol.} \vo{149} \pp{381}{424}.

De Boer, R. J., A. U. Neumann, A. S. Perelson, L. A. Segel and
G. Weisbuch. 1992a. \ti{Recent approaches to immune networks}
\in{Proc. First European
Biomathematics Conf.}, V. Capasso and P. Demongeot, eds.,
Springer-Verlag: Berlin (in press).

De Boer, R. J., L. A. Segel and A. S. Perelson. 1992b.
\ti{Pattern formation in one and two-dimensional shape space
models of the immune system} \jo{J. Theor. Biol.} \vo{155}
\pp{295}{233}.

De Boer, R. J., I. G. Kevrekidis and A. S. Perelson. 1993a.
\ti{Immune network behavior I: From stationary states
to limit cycle oscillations}
\jo{Bull. Math. Biol.} (in press).

De Boer, R. J., I. G. Kevrekidis and A. S. Perelson. 1993b.
\ti{Immune network behavior II: From oscillations to chaos and
stationary states} \jo{Bull. Math. Biol.} (in press).

Doedel, E. J. 1981.
\ti{AUTO: a program for the bifurcation analysis of
autonomous systems}
\jo{Cong. Num.} \vo{30} 265-285.

Farmer, J. D., N. H. Packard and A. S. Perelson. 1986.
\ti{The immune system, adaptation, and machine learning}
\jo{Physica} \vo{22D} \pp{187}{204}.

Fortmann, T. E. and K. L. Hitz. 1977
\bt{An Introduction to Linear Control Systems}, p. 170,
Marcel Dekker, Inc.: New York.

Hiernaux, J. R., P. J. Baker and C. DeLisi. 1982.
\ti{Oscillatory immune response to lipopolysaccharide}
\in{Regulation of Immune Response Dynamics, Vol. 1}
C. DeLisi and J. Hiernaux, eds.,
\pp{121}{136}
CRC Press: Boca Raton, FL

Hoffmann, G. W. 1975.
\ti{A theory of regulation and self-nonself discrimination in
an immune network} \jo{Eur. J. Immunol.} \vo{5} \pp{638}{647}.

Hoffmann, G. W. 1982.
\ti{The application of stability criteria in evaluating
network regulation models}
\in{Regulation of Immune Response Dynamics, Vol. 1}
C. DeLisi and J. Hiernaux, eds., \pp{137}{162}
CRC Press: Boca Raton, FL

Holmberg, D., S. Forsgen, F. Ivars and A. Coutinho. 1984.
\ti{Reactions among IgM antibodies derived from normal
neonatal mice}
\jo{Eur. J. Immunol.} \vo{14} 435--441.

Holmberg, D. 1987.
\ti{High connectivity, natural antibodies preferentially
use 7183 and QUPC52 V-H families}
\jo{Eur. J. Immunol.} \vo{17} 399--403.

Holmberg, D.,  \AA. Andersson, L. Carlson and S. Forsgen. 1989.
\ti{Establishment and functional implications of B-cell
connectivity}
\jo{Immunol. Rev.} \vo{110} \pp{89}{103}.

Hooykaas, H., R. Benner, J. R. Pleasants and B. S. Wostmann.  1984.
\ti{Isotypes and specificities of immunoglobulins produced by
germ-free mice fed chemically defined ``antigen-free" diet}
\jo{Eur. J. Immunol.} \vo{14} \pp{1127}{1130}.

Huang, J.-Y., R. E. Ward and H. K\"ohler. 1988.
\ti{Biological mimicry of antigenic stimulation: analysis of the
in vivo antibody responses induced by monoclonal anti-idiotypic
antibodies}
\jo{Immunol.} \vo{63} pp.1-8.

Jerne, N. K. 1974.
\ti{Towards a network theory of the immune system}
\jo{Ann. Immunol. (Inst. Pasteur)} \vo{125 C} \pp{373}{389}.

Jerne, N. K., J. Roland and P. A. Cazenave. 1982.
\ti{Recurrent idiotypes and internal images}
\jo{EMBO J.} \vo{1} 243--247.

Kearney, J. F. and M. Vakil. 1986.
\ti{Idiotype-directed interactions during ontogeny play a major
role in the establishment of the B cell repertoire}
\jo{Immunol. Rev.} \vo{94} 39--50.

Kearney, J. F., M. Vakil and N. Nicholson. 1987.
\ti{Non-random VH gene expression and idiotype anti-idiotype
expression in early B cells}
\in{Evolution and Vertebrate Immunity: the antigen Receptor
and MHC Gene Families}
G. Kelsoe and D. Schulze, eds.,
\pp{373}{389} Austin: University of Texas Press.

K\"ohler, H., H. L. Cheng, A. K. Sood, M. McNamara-Ward,
J. H. Huang, R. E. Ward and
T. Kieber-Emmons. 1986.
\ti{On the mechanism of internal image vaccines}
In: {\it Idiotypes}, M. Reichlin and J. D. Capra, eds.,
Orlando: Academic Press
\pp{179}{190}.

K\"ohler, H., T. Kieber-Emmons, S. Srinivasan, S. Kaveri,
W. J. W. Morrow, S. Muller, C. Kang and
S. Raychaudhuri. 1989.
\ti{Revised immune network concepts}
\jo{Clin. Immunol. Immunopathol.}
\vo{52} pp. 104.

K\"ohler, H. 1991.
\ti{The servant idiotype network}
\jo{Scand. J. of Immunol.} \vo{33} 495--497.

Lundkvist, I., A. Coutinho, F. Varela and D. Holmberg. 1989.
\ti{Evidence for a functional idiotypic network amongst
natural antibodies in normal mice}
\jo{Proc. Nat. Acad. Sci. USA.} \vo{86} \pp{5074}{5078}.

Neumann, A. and G. Weisbuch. 1992a.
\ti{Window automata analysis of population dynamics in
the immune system}
{\it Bull. Math. Biol.} \vo{54} 21--44.

Neumann, A. U. and G. Weisbuch. 1992b.
\ti{Dynamics and topology of immune networks}
\jo{Bull. Math. Biol.} \vo{54} 699--726.

Novotn\'y, J., M. Handschumacher, and R. E. Bruccoleri. 1987.
\ti{Protein antigenicity: a static surface property}
\jo{Immunol. Today} \vo{8} 26--31.

Pereira, P., L. Forni, E. L. Larsson, M. Cooper, C.
Heusser and A. Coutinho. 1986.
\ti{Autonomous activation of B and T cells in antigen-free mice}
\jo{Eur. J. Immunol} \vo{16} \pp{685}{688}.

Perelson, A. S. 1984. \ti{Some mathematical models of
receptor clustering by multivalent antigens}
\in{Cell Surface Dynamics:  Concepts
and Models}, A. S. Perelson, C. DeLisi and F. W.
Wiegel, Eds., \bp{223}{276}.  Marcel Dekker: New York.

Perelson, A. S. 1989.
\ti{Immune network theory} \jo{Immunol. Rev.} \vo{110} \pp{5}{36}.

Perelson, A. S. and C. DeLisi. 1980.
\ti{Receptor clustering on a cell surface. I. Theory of
receptor cross-linking by ligands bearing two chemically
identical functional groups}
\jo{Math. Biosci.} \vo{48} \pp{71}{110}.

Perelson, A. S. and G. Weisbuch. 1992.
\ti{Modeling immune reactivity in secondary lymphoid organs}
\jo{Bull. Math. Biol.} \vo{54} \pp{649}{672}.

Raychaudhuri, S., C. Kang, S. Kaveri,
T. Kieber-Emmons and H. K\"ohler. 1990.
Tumor idiotype vaccines VII. Analysis and correlation of structural,
idiotypic, and biological properties of protective and
nonprotective Ab2.
\jo{J. Immunol.} \vo{145} pp. 760-767.

Richter, P. H. 1975
\ti{A network theory of the immune system}
\jo{Eur. J. Immunol.} \vo{5} \pp{350}{354}.

Romball, C. G. and W. O. Weigle. 1982.
\ti{Cyclic antibody production in immune regulation}
\in{Regulation of Immune Response Dynamics, Vol. 1}
C. DeLisi, and J.  Hiernaux, eds., \pp{9}{26}
CRC Press: Boca Raton, FL

Segel L. A. and A. S. Perelson. 1989.
\ti{Shape space analysis of immune networks}
\in{Cell to Cell Signaling:
{}From Experiments to Theoretical Models}, (A. Goldbeter, ed.),
\bp{273}{283}, Academic Press: New York.

Segel L. A. and A. S. Perelson. 1991.  \ti{Exploiting the
diversity of time scales
in the immune system: A B-cell antibody model}
\jo{J. Stat. Phys.} {\bf 63}, 1113--1131.

Segel, L. A. and A. S. Perelson. 1988.
\ti{Computations in shape space.  A new approach to
immune network theory}
\in{Theoretical Immunology, Part Two, SFI Studies in the
Sciences of Complexity},
A. S. Perelson, ed.,
Addison-Wesley: Redwood City, CA, pp. 321--343.

Segel, L. A. and A. S. Perelson. 1989.
\ti{Shape space analysis of immune networks}
\in{Cell to Cell Signaling: From Experiments to Theoretical Models}
A. Goldbeter, ed., Academic Press: NY, pp. 273--283.

Sparrow, C. 1982.
\ti{The Lorenz equations: bifurcations, chaos and strange attractors}
\jo{Applied Mathematical Sciences}
\vo{41} Springer-Verlag: New York.

Spouge, J. L. 1986.
\ti{Increasing stability with complexity in a system
composed of unstable subsystems}
\jo{J. Math. Analysis and Applications}
\vo{118} \pp{502}{518}

Stewart, J. and F. J. Varela. 1989.
\ti{Exploring the meaning of connectivity in the immune network}
\jo{Immunol. Rev.} \vo{110} \pp{37}{61}.

Stewart, J. and F. J. Varela. 1990.
\ti{Dynamics of a class of immune networks. II. Oscillatory
activity of cellular and humoral components}
\jo{J. Theor. Biol.} \vo{144} \pp{103}{115}.

Stewart, J. and F. J. Varela. 1991.
\ti{Morphogenesis in shape-space: Elementary meta-dynamics
in a model of the immune network}
\jo{J. Theoret. Biol.} \vo{154} 477--498.

Taylor, M. A. and I. G. Kevrekidis. 1990.
\ti{Interactive AUTO: A graphical interface for AUTO86}
Technical Report, Dept. of Chemical Engineering, Princeton
University.

Varela, F. J. and A. Coutinho. 1991. \ti{Second generation
immune networks} \jo{Immunol. Today} \vo{12} \pp{159}{166}.

Varela, F., A. Andersson, G. Dietrich, A. Sundblad,
D. Holmberg, M. Kazat\-ch\-kine and A. Coutinho. 1991.
\ti{The population dynamics of natural antibodies in normal and
autoimmune individuals} \jo{Proc. Nat. Acad. Sci. USA}
\vo{88} \pp{5917}{5921}.

Weigle, W. O. 1975.
\ti{Cyclical production of antibody as a regulatory mechanism
in the immune response}
\jo{Adv. Immunol.}
\vo{21} \pp{87}{111}.

Weinand, R. G. 1990.
\ti{Somatic mutation, affinity maturation and the antibody
repertoire: a computer model}
\jo{J. Theoret. Biol.} \vo{143} |pp{343}{382}.

Weisbuch, G. 1990.
\ti{A shape space approach to the dynamics of the immune system}
\jo{J. Theoret. Biol.} \vo{143} pp. 507-522.

Weisbuch, G., R. J. De Boer and A. S. Perelson. 1990.
\ti{Localized memories in idiotypic networks}
\jo{J. Theoret. Biol.} \vo{146} \pp{483}{499}.

\vfill\eject

\skip
\leftline{\bigbf FIGURE CAPTIONS}

\skip
\skip
Figure 1.  Topology of a homogeneous Cayley tree.
Each node represents a clone -- both the B cell population
and its secreted antibody concentration.
Each clone is connected to $z$ adjacent clones.
A Cayley tree with coordination number $z=1$ is
equivalent to a two-clone model.
A Cayley tree with
$z=2$ corresponds to a linear chain with clone 1 as the root
of the tree. With $z\geq 3$, a Cayley tree is a representation
of a network without loops.

\skip
\skip
Figure 2. Dynamical response to a perturbation of a localized
tolerance attractor at level 2.
A perturbation of the localized steady-state
at  level 2 returns to its attractor.
(a) B cells and (b) antibodies at levels 1 through 5 are shown.
The connectivity parameter, $z$, is set to 16,
where a localized memory cannot exist.
Other system parameters are set to their standard values:
$\theta_1=100$, $\theta_2=10^4$, $p=1$, $s=1$, $m=1$,
$d_B=0.5$, $d_A=0.05$, $d_C=0.01$. The initial conditions
are
$b_1=5000$, $a_1=9000$, $b_2=17000$, $a_2=5000$,
$b_3=631$, $a_3=3.19$, $b_4=3.16$, $a_4=2.1$,
$b_5=2.09$, $a_5=0.605$,
$b_6 - b_{10}=2$, $a_6 - a_{10} =0$.

\skip
\skip
Figure 3. Bifurcation diagram with $z$ as the
bifurcation parameter.
All other parameters are set to
their standard values (see text).
The vertical axis indicates the
highest B cell population in the localized
state (i.e. $B_1$ for the immune state;
$B_2$ for the tolerant state).
The localized steady-state remains stable for a wide
range of values for $z$.
The lower branch is unstable.
The localized state at
level 2 exists for a slightly larger range of $z$ than the
localized state at level 1.
Steady states also exist for
larger values of $z$, but they correspond to ``extended localization"
attractors or ``percolation
attractors", where clones at many levels are sustained at
high steady-state populations.
(See Fig. 5)

\skip
\skip
Figure 4. Nondimensional level 3 field versus $z$.
The field experienced by level 3 clones,
$H_3=A_2+(z-1)A_4$, consists of two components.
Steady state estimates in Section 3
were based on the assumption that
$A_2$ dominates the field.
As $z$ is increased, however, this assumption breaks
down, and the localized immune state is lost.

\skip
\skip
Figure 5. Extended localized attractor in high $z$.
Example of a system attractor past the limit point for
a localized memory state ($z=16$).
Activation at many levels is referred to a ``extended
localized state"
(Neumann and Weisbuch, 1992a).
At $t=0$ a system in a localized immune state for $z=15$
has $z$ increased to 16. This localized state is slowly
lost, and after a transient, an extended localized
attractor is attained.
When $z$ is increased further, this extended localization
breaks down, and the system converges on a percolation attractor.
The nondimensional concentrations of  (a) B cells and (b)
 antibodies at levels 1 through 5 are shown.
Other system parameters are set to their standard values:
$\Theta_1=0.1$, $\Theta_2=10$,
 $\delta=0.1$, $\sigma=0.04$, $\rho=2$, $ \nu=0.1$,
$\mu = 20$, $\alpha= 1000$, and $\beta= 50$.
The initial conditions are
$B_1=418$, $A_1=9.773$, $B_2=26.8$, $A_2=0.0068$,
$B_3=0.134$, $A_3=0.00396$, $B_4=0.112$, $A_4=0.0034$,
$B_5=0.10$, $A_5=0.0031$,
$B_6=0.089$, $A_6=0.0028$,
$B_7=0.077$, $A_7=0.0025$,
$B_8=0.062$, $A_8=0.0021$,
$B_9=0.047$, $A_9=0.0013$,
$B_10=0.041$, $A_10=0.00043$.

\skip
\skip
Figure 6. Bifurcation diagram of the localized immune
state with $\delta$ as a variable.
All other parameters are set to their standard values.
The nondimensional $B_1$ population is plotted.
As $\delta$ drops below 0.0136, a Hopf bifurcation occurs.
The branch of the Hopf bifurcation
consists of unstable limit cycles,
while continuation of the primary branch leads
to an unstable steady-state.
Most attractors in this region appear to be chaotic.

\skip
\skip
Figure 7. Bifurcation diagram of the localized immune state
with $\mu$ as a variable.
The solid line indicates the nondimensional $B_1$
population in the localized, stable immune
steady-state.
Continuation through a Hopf bifurcation (at $\mu=12.45$)
leads to
an unstable steady-state with 2 unstable complex eigenvalues.
(The numbers in the figure legend indicate the number of
eigenvalues with a positive real part on each branch).
After the saddle-node bifurcation,
an additional real positive eigenvalue
appears, which gets larger for larger values of $\mu$.
A second Hopf bifurcation occurs at $\mu=12.1$
as the complex eigenvalues re-cross the imaginary axis, but
the single positive eigenvalue persists.
Both branches born at the Hopf bifurcations define
unstable limit cycles.

\skip
\skip
Figure 8. A two parameter ($\mu ,\delta$)
continuation of the localized immune state.
Assuming the standard parameter set for all other
values, this diagram shows the combinations of
$\mu$ and $\delta$ for which the localized memory
state at level 1 exists as well as whether it is stable.
Legend Key: SN = Saddle-node, HB-1,2 = Hopf bifurcation
curves.
The localized steady-state does not exist in the parameter
regime below the saddle-node curve.
The localized steady-state is stable only above the
upper Hopf bifurcation curve (HB-1).
This diagram qualitatively corresponds to Fig. 3 in
Perelson and Weisbuch (1992).

\skip
\skip
Figure 9. Two parameter continuation ($z ,\mu$)
of the localized immune state.
All other parameters set to the standard values.
Network connectivity, $z$, determines the existence
of the localized steady-state; while the dynamical
parameter, $\mu$, determines the stability.
The steady-state is unstable below the upper Hopf bifurcation
curve.

\skip
\skip
Figure 10. Time plots and
phase plot projections of attractors as the
localized steady-state becomes unstable.
A 2-dimensional projection of a 10-dimensional state-space
into the $a_1-a_2$ plane is shown.
Because this is a projection, trajectories may cross.
Parameter values:
(a, b) $d_C=0.01$,
(c, d) $d_C=0.0067$,
(e, f) $d_C=0.0060$, and
(g, h) $d_C=0.0010$.
Other parameters are set to their standard values.
The initial conditions  are
$b_1=13900$, $a_1=9800$, $b_2=1300$, $a_2=10$,
$b_3=3.4$, $a_3=3.2$, $b_4=2.4$, $a_4=1.7$,
$b_5=2.07$, $a_5=0.58$,
$b_6 - b_{10}=2$, $a_6 - a_{10} =0$.
B cells and antibodies at levels 1 through 5 are denoted
by the symbols
asterisk, box, octagon, diamond and cross, respectively.

\skip
\skip
Figure 11. Dynamics of a chaotic attractor.
The parameter $d_C$ is set past the Hopf bifurcation
($d_C=.005$), and hence in the chaotic regime.
At time $t=100$, the system is perturbed
by a large ($10^5$) dose of $Ab_1$.
(a) $z=2$, (b) z=33.
The same sized
injection has little effect on the more highly connected
network.
Other system parameters are set to their standard values:
$\theta_1=100$, $\theta_2=10^4$, $p=1$, $s=1$, $m=1$,
$d_B=0.5$, $d_A=0.05$.
The initial conditions are
$b_1=13900$, $a_1=9800$, $b_2=1300$, $a_2=10$,
$b_3=3.4$, $a_3=3.2$, $b_4=2.4$, $a_4=1.7$,
$b_5=2.07$, $a_5=0.58$,
$b_6 - b_{10}=2$, $a_6 - a_{10} =0$.

\skip
\skip
Figure 12. Phase and time plots of a localized limit cycle (a,b),
a localized oscillatory transient (c,d),
and a localized chaotic attractor (e,f).
Attractors in a highly connected network ($z=100$)
with small bone marrow source term ($m=2.5 X 10^{-5}$)
do not necessarily activate levels deeper in the network
simply due to oscillatory behavior.
All three trajectories shown orbit two unstable
steady states.
Initial conditions for limit cycle ($d_C=.00009$):
$b_1=46.3$, $a_1=30.7$, $b_2=.165$, $a_2=75.5$,
$b_3=.000785$, $a_3=.00393$,
$b_4 - b_{10}= 10^{-6}$, $a_4 - a_{10} =0$.
Initial conditions for localized transient
($d_C = .0025$) and localized chaos ($d_C=.005$):
$b_1=7270$, $a_1=6970$, $b_2=126$, $a_2=2.04$,
$b_3 - b_{10}= .00005$, $a_3 - a_{10} =0$.
Other system parameters:
$\theta_1=100$, $\theta_2=10^4$, $p=1$, $s=1$,
$d_B=0.5$, $d_A=0.05$.

\skip
\skip
Figure 13. Schematic diagram of some
idiotypic interactions
absent in a Cayley tree model.
Internal images ($Ab_{2\beta}$) mimic the structure of the
original antigenic epitope (Ag); therefore, they are
topologically substitutable for antigen in a network model.
$Ab_{2\alpha}$'s, which do not
mimic antigenic structure, yet crossreact with more than
one $Ab_1$ may serve as a model for network antigens.

\bye